\documentclass[a4paper]{webofc}
\usepackage[varg]{txfonts}   
\wocname{ICNFP 2018}
\woctitle{ICNFP 2018}
\begin{document}
\selectlanguage{english}
\title{Exotic Baryons in Chiral Soliton Models}
%
%

\author{H. Weigel\inst{1}
}

\institute{Institute for Theoretical Physics, Physics Department, \\
Stellenbosch University, Matieland 7602, South Africa}

\abstract{%
We cautiously review the treatment of pentaquark exotic baryons in chiral soliton models.
We consider two examples and argue that any consistent and self-contained description must go beyond
the mean field approximation that only considers the classical soliton and the canonical quantization
of its (would-be) zero modes via collective coordinates.
}
\maketitle

\section{Introduction}
Exotic baryons carry quantum numbers that cannot be constructed as products of three quark states. 
In this discussion, which is based on Refs.~\cite{Walliser:2005pi,Weigel:2007yx,Blanckenberg:2015dsa}, 
we will examine and review their description in the framework of chiral soliton models; see 
Ref.~\cite{Weigel:2008zz} for a recent review on these models. We distinguish between two types 
of exotic baryons; those with only the three light flavors (up, down, strange) and those that contain 
heavy flavors like charm and bottom. Though their description in soliton models is fundamentally different, 
both require a three flavor quantization of the soliton and considerations beyond the mean field approach.

It is important to stress that exotic baryons are actually resonances. Hence they must be observed 
in scattering processes. To put this into the context of chiral soliton models we first investigate 
pentaquarks with light flavors and their role in the Skyrme model description of kaon nucleon
scattering. By definition of solitons, no term linear in meson fluctuations, that eventually could be
identified as a Yukawa interaction, should emerge. If it does, it is a mere short-coming of approximating
the exact time-dependent soliton solution. Rather, in scattering the resonance content must be directly
analyzed to obtain the widths of (collective) resonances. This implies that the decay width may not be
estimated from axial current matrix elements of the collectively excited classical soliton. In turn, 
this suggests that mean field approaches as {\it e.g.} in 
Refs.~\cite{Diakonov:1997mm,Ellis:2004uz,Kim:2017khv,Praszalowicz:2018upb} are not appropriate to 
describe resonance widths for hadronic decays. The full calculation indeed requires the introduction 
of both, collective and harmonic excitations of the classical soliton. The canonical formulation of 
these fluctuations also induces the constraints needed to avoid double counting errors. An alternative 
approach that approximates the collective fluctuations to be harmonic is known to be exact in the 
academic limit of infinitely many color degrees of freedom ($N_C\to\infty$). Since our approach may 
be applied to any (odd) value $N_C$ we compare and indeed find identical scattering data when
$N_C\to\infty$.

The second example deals with pentaquarks that contain a single heavy quark (charm or bottom) or 
anti-quark. The central element of heavy quark physics is the corresponding spin-flavor symmetry that 
predicts a degeneracy of heavy pseudoscalar and vector mesons~\cite{Eichten:1980mw,Neubert:1993mb}. A 
model with only chiral fields in a mean field formulation~\cite{Yang:2016qdz}\footnote{Listing the 
references on the mean field description for heavy baryons in soliton models can hardly be 
exhaustive. The interested reader my trace them from the review~\cite{Kim:2018cxv}.} will not be 
sufficient because it cannot combine these fields in a single multiplet. The model must be augmented 
accordingly with these additional fields and couple them to the chiral soliton. This advances the bound 
state approach to strangeness~\cite{Callan:1985hy} and particularly shows that the heavy meson bound 
state selects the appropriate $SU(3)$ representation for the light flavor (up, down, strange) component 
of the baryon wave-function which models the light diquark structure. Basic heavy baryons select the 
anti-triplet and the sextet. Pentaquarks with a heavy anti-quark relate to the anti-sextet while those 
with a light anti-quark have light flavors from the anti-decapentaplet. 

\section{Chiral soliton models}
\label{exotic}
Chiral soltion models are low-energy models for baryons. They are based on the conjecture, resulting 
from large $N_C$ considerations of QCD, that baryons are described as soliton solutions of a non-linear 
effective meson theory~\cite{Witten:1979kh}. At low energies chiral symmetry is governing this effective 
theory. The most prominent such model is the Skyrme model~\cite{Skyrme:1961vq}, which was firstly applied 
to static baryon properties in Ref.~\cite{Adkins:1983ya}. That model only contains the pseudoscalar mesons 
and is too simple for a detailed description of baryon properties.\footnote{For example, other meson 
fields are needed to reproduce the neutron-proton mass difference~\cite{Jain:1989kn}. Consult the 
review~\cite{Weigel:2008zz} for further examples.} Here we will nevertheless base our arguments on
this model because it nicely illuminates the structure of chiral soliton models: In most aspects the 
generation of baryons states via canonical quantization of the soliton's collective motion is the crucial 
ingredient while the microscopic details are of lesser relevance.

We are dealing with three light degrees of freedom so that the effective meson theory is a 
functional of the chiral field $U(\vec{x},t)\in SU(3)$. Except for flavor symmetry breaking,
the theory is invariant under global chiral transformations $U\to LUR^\dagger$ while symmetry 
breaking terms transform as the eighth component of an octet of the vector subgroup $L=R$.
The Skyrme soliton itself is a mapping from configuration space to $SU(2)$. In the presence 
of flavor symmetry breaking it must be embedded in the isospin subgroup to minimize the classical 
energy. Then the static soliton assumes the {\it hedgehog} structure
\begin{equation}
U_H(\vec{x}\,)=\begin{pmatrix}
\begin{array}{c|c}
\hspace{0.1cm}{\rm exp}\left(i\hat{\vec{x}} \cdot \vec{\tau\,} F(r)\right)
\hspace{0.1cm} & \hspace{0.3cm}
\begin{picture}(0,0)
\put(-0.1,-0.15){\mbox{\footnotesize $0$}}
\put(-0.1,0.15){\mbox{\footnotesize $0$}}
\end{picture}\hspace{0.3cm} \cr
\hline
0\hspace{1.0cm}0 & 1
\end{array}\end{pmatrix}\,.
\label{eq:hedgehog}\end{equation}
The profile function $F(r)$ (called chiral angle) is obtained from the variational
principle to the classical energy, $E_{\rm cl}$. The boundary condition $F(0)=\pi$ and 
$\lim_{r\to\infty}F(r)=0$ enforces unit topological charge, which is identified with the 
baryon number~\cite{Wi83}. The classical energy only plays a minor role in the soliton 
picture of baryons because the soliton model is not renormalizable so that quantum 
corrections (the vacuum polarization energy) are difficult to control~\cite{Meier:1996ng}. 
Hence we focus on baryon mass differences in which both the classical and the vacuum 
polarization energies cancel.

In the next step towards generating baryon states, time dependent solutions must be
obtained. These are not known but there are two major \underline{approximations} in three 
flavor models. One is based on the so-called bound state approach~\cite{Callan:1985hy} that 
introduces harmonic fluctuations in strangeness direction. We will get back to this approach 
in the context of heavy baryons. The other approximation introduces time dependent collective 
coordinates for the (approximate) vector symmetry~\cite{Guadagnini:1983uv}
\begin{equation}
U(\vec{x},t)=A(t)U_H(\vec{x}\,)A^\dagger(t)
\qquad {\rm with}\qquad A(t)\in SU(3)\,,
\label{eq:coll1}\end{equation}
thereby generalizing the two flavor procedure of Ref.~\cite{Adkins:1983ya}. Integrating
over space yields the collective coordinate Lagrangian (flavor symmetry breaking excluded,
for the time being)
\begin{equation}
L=-E_{\rm cl}[F]+\frac{1}{2}\alpha^2[F]\sum_{i=1}^3\Omega_i^2
+\frac{1}{2}\beta^2[F]\sum_{\alpha=4}^7\Omega_\alpha^2
-\frac{N_C}{2\sqrt{3}}\,\Omega_8
\qquad {\rm where}\qquad 
\Omega_a=-{\rm i}{\rm tr}\left[\lambda_aA^\dagger(t)\,\frac{dA(t)}{dt}\right]\,.
\label{eq:coll2}\end{equation}
The moment of inertia tensor separates into non-strange~($\alpha^2$) and strange~($\beta^2$) 
components due to the particular embedding, Eq.~(\ref{eq:hedgehog}). These are functionals
of the soliton and the actual value depends on the particular model though they are always 
$\mathcal{O}(N_C)$. The last term is only linear in the time derivative with a fixed coefficient. 
In purely mesonic theories it arises from the Wess-Zumino term~\cite{Jain:1984gp} while in 
chiral quark soliton models it stems from the interaction of the chiral field with 
self-consistently constructed quark fields~\cite{Alkofer:1994ph}. 

\section{Collective coordinate quantization with three flavors}
\label{SU3qaunt}
The generation of baryon states proceeds by canonically quantizing the angular 
degrees of freedom in $A(t)$. This introduces eight right $SU(3)$ generators
\begin{equation}
R_a=-\frac{\partial L}{\partial\Omega_a}=\begin{cases}
-\alpha^2\Omega_a=-J_a,&a=1,2,3\cr
-\beta^2\Omega_a,&a=4,..,7\cr
\frac{N_C}{2\sqrt3},&a=8\,,
\end{cases}
\label{eq:Rgen}\end{equation}
that are subject to the commutation relations
$\left[R_a,R_b\right]=-if_{abc}R_c$.
A Legendre transformation yields the Hamiltonian for the collective coordinates
\begin{align}
H-E_{\rm cl}=\frac{1}{2\alpha^2}\vec{J\,}^2
+\frac{1}{2\beta^2}\sum_{\alpha=4}^7R_\alpha^2+H_{\rm SB}(A)
=\frac{1}{2}\left(\frac{1}{\alpha^2}-\frac{1}{\beta^2}\right)\vec{J\,}^2
+\frac{1}{2\beta^2}C_2[SU(3)]-\frac{N_C^2}{24\beta^2}+H_{\rm SB}(A)\,,
\label{ch6:hamrigid1}
\end{align}
where $C_2[SU(2)]=\sum_{a=1}^8R_a^2$ is the quadratic Casimir operator of $SU(3)$. 
Again flavor symmetry breaking terms have not been made explicit in $H_{\rm SB}(A)$.
We diagonalize this Hamiltonian subject to the constraint
$Y_R=\frac{2}{\sqrt{3}}R_8=\frac{N_C}{3}$. Since $\left[H,R_8\right]=0$ this can 
be directly imposed on the eigenstates. Without symmetry breaking these are 
elements of $SU(3)$ representations. 

\begin{figure}
\centerline{
\includegraphics[width=10cm,height=4cm]{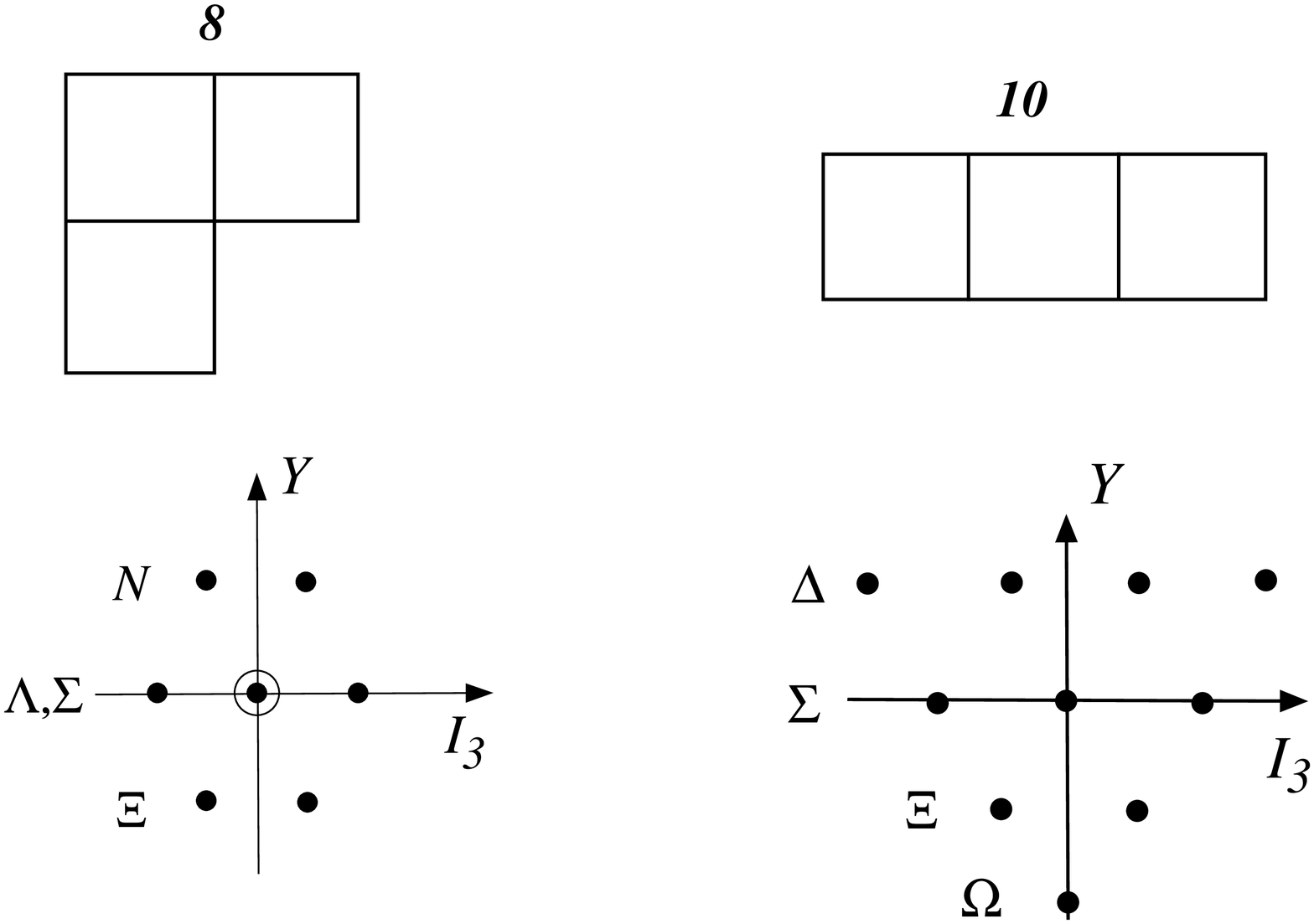}}
\smallskip
\caption{\label{fig:ground}Young tableaux and particle content of the 
low-lying $SU(3)$ representations.}
\end{figure}

For $N_C=3$ we have $Y_R=1$ and the representations with the lowest $C_2$ eigenvalues 
are the octet ($J=\frac{1}{2}$) and the decuplet ($J=\frac{3}{2}$) \cite{Guadagnini:1983uv}.
Their Young tableaux and particle content are shown in Fig.~\ref{fig:ground}.
In Young tableaux a single box represents a quark while the adjoint,
\parbox[l]{3mm}{\vskip-1.0mm\includegraphics[width=2mm,height=4mm]{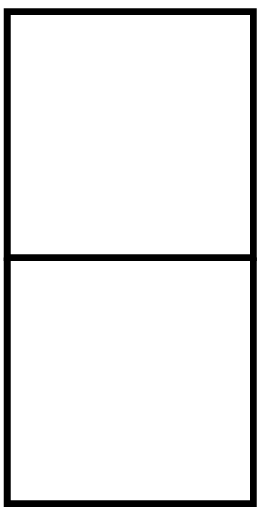}}
stands for either two quarks or an antiquark, depending on the prescribed baryon number. 
Hence, for unit baryon number, the octet and decuplet represent three quark composites.

\begin{figure}
\centerline{
\includegraphics[width=10cm,height=4cm]{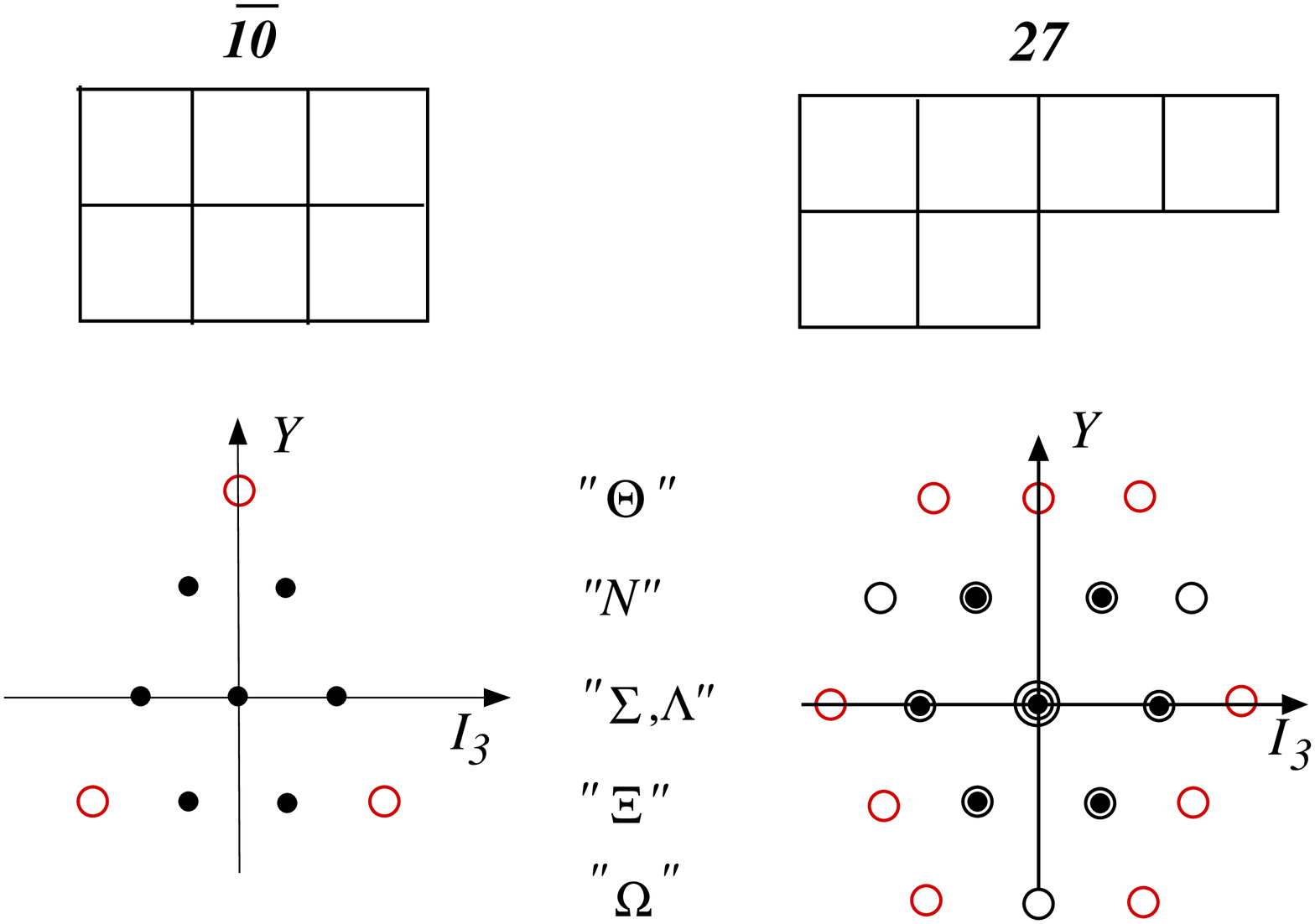}}
\smallskip
\caption{\label{fig:higher}(Color online) Young tableaux and particle content the 
anti-decuplet and the 27-plet $SU(3)$ representations. Red circles indicate
exotic baryons that cannot be built from three quarks.}
\end{figure}

However, higher dimensional representations also solve the (flavor symmetric) eigenvalue
problem. Examples are the anti-decuplet and the 27-plet that are shown in Fig.~\ref{fig:higher}.
The Young tableaux interpretation immediately reveals that these higher dimensional
representations contain baryons with additional quark anti-quark pairs. Even more, some
baryons in these representations have quantum numbers corresponding to exotic, non-three 
quark states\footnote{To the author's knowledge the occurrence of these exotic baryons  
in the collective quantization scheme was first noticed by Biedenharn and Dothan
\cite{Biedenharn:1984su}.}. The most prominent example is the $\Theta^{+}$ pentaquark 
in the anti-decuplet whose valence quark content is $\overline{s}uudd$. The
$C_2$ eigenvalues of the octet and the anti-decuplet are 3 and 6, respectively. Hence,
in a flavor symmetric world the mass difference between the $\Theta^{+}$ and the nucleon
would be $3/2\beta^2$ (for $N_C=3$). 

These higher dimensional representations play another important role when including
$SU(3)$-flavor symmetry breaking that is reflected by the hadron masses varying with 
their strangeness content. The eigenstates of the collective coordinate Hamiltonian
are no longer elements of a specific $SU(3)$ representation. Rather, baryon states 
with identical quantum numbers from different representations mix as the perturbative
diagonalization of $C_2+2\beta^2 H_{\rm SB}$ nicely reveals \cite{Park:1989by}. Even
more, as pointed out by Yabu and Ando \cite{Yabu:1987hm} this operator can indeed 
be diagonalized exactly (numerically) and higher order effects may be relevant in 
particular cases.  We will get back to this in the context of heavy baryons.

\section{$\Theta$ mass and width}
\label{theta}
The essential question is: Do exotic elements of higher dimensional representations stand 
for sensible resonances, or are they merely artefacts of the collective coordinate 
quantization? This issue arises already in the simpler two flavor version that predicts 
an exotic spin $J=\frac{5}{2}$ state well below $2{\rm GeV}$~\cite{Dorey:1994fk}.

To answer this question for soliton model calculations, it is imperative to recall 
the meaning of resonances in potential scattering. In the particular case of 
single channel potential scattering a resonance is signaled by the transition of the 
phase shift through $\frac{\pi}{2}$ with a positive slope \cite{Newton:1982qc}. 
To identify the potential, time dependent harmonic fluctuations about the classical
soliton, Eq.~(\ref{eq:hedgehog}) are introduced~\cite{Schwesinger:1988af}. 
When augmenting the rotating hedgehog, Eq.~(\ref{eq:coll1}) 
by time dependent fluctuations linear terms emerge that couple to the collective rotations. 
In early soliton model studies these terms were identified as Yukawa interactions and their 
transition matrix elements were used to estimate widths of $\Delta$-resonances~\cite{Adkins:1983ya}. 
An alternative to compute a width from a transition matrix element was to identify the pion field 
with axial current of the rotating soliton. This was justified as a generalized Goldberger-Treiman 
relation and was applied to the $\Theta^{+}$ suggesting a surprisingly narrow 
resonance~\cite{Diakonov:1997mm,Ellis:2004uz}. Yet this generalization is questionable 
since the soliton field equation is the same for $SU(2)$ and $SU(3)$ but the axial currents 
are not~\cite{Weigel:2007yx}. As a matter of fact, additional $1/N_C$ contributions to 
the axial current are not accounted for in PCAC. Neither does the soliton part of the chiral 
field correspond to an asymptotic meson state.

In any event, by pure definition of the soliton as a stationary point of the action there 
cannot exist a term linear in small amplitude fluctuations. If it nevertheless emerges, it 
just reflects that the background configuration is at best an approximation to the actual 
soliton. Thus there is natural scepticism on plain Yukawa interactions in soliton models. 
This was definitely recognized early on as there have been numerous attempts \cite{width}
to improve on the $\Delta$-width estimate of Ref.~\cite{Adkins:1983ya}.

\begin{figure}
\centerline{
\includegraphics[width=7.0cm,height=4.5cm]{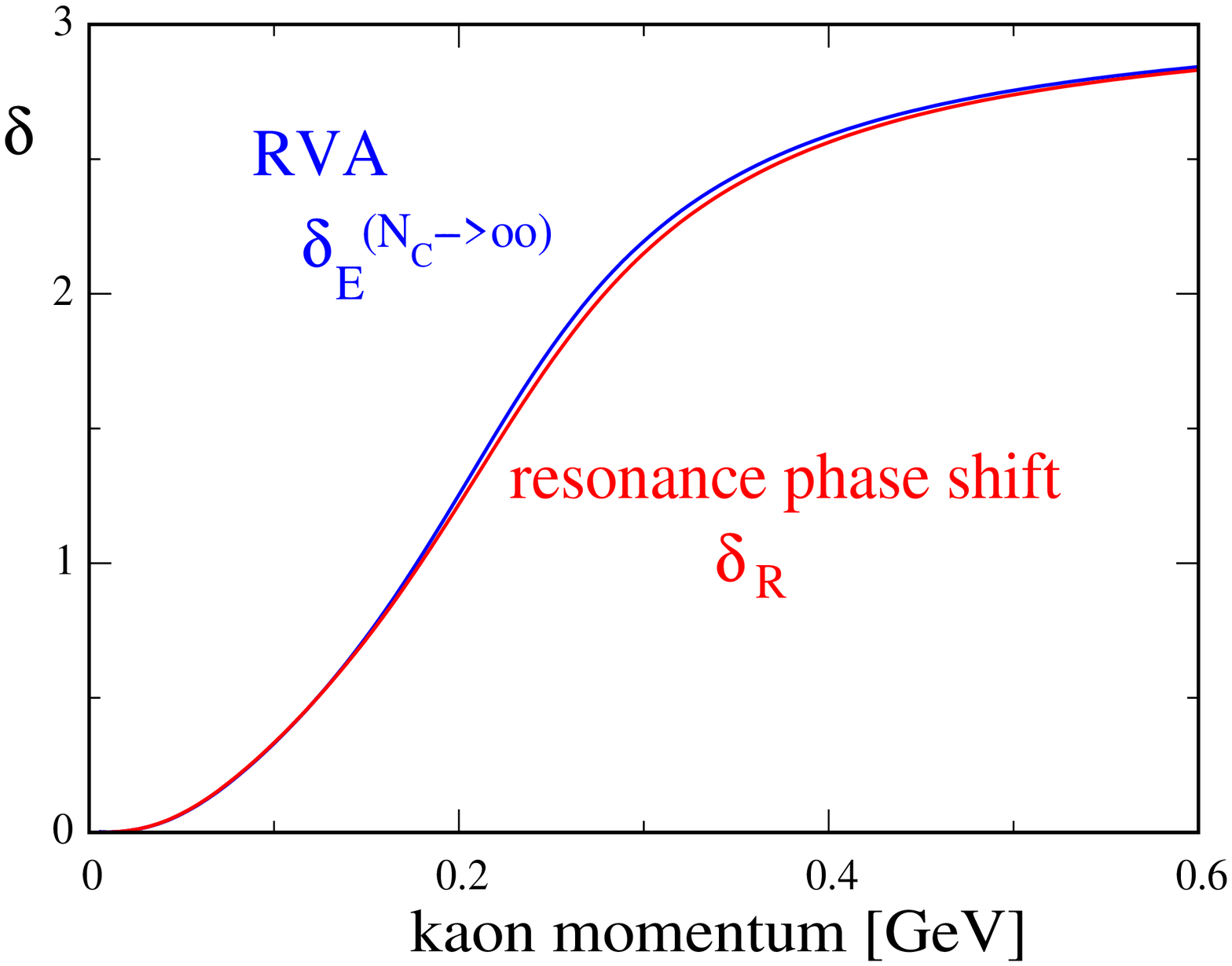}\hspace{5mm}
\includegraphics[width=7.0cm,height=4.5cm]{width.eps}}
\caption{\label{fig:delta}(Color online) Skyrme model results for phase shifts and decay
widths of kaon nucleon scattering in the $\Theta^{+}$ channel.}
\end{figure}

When examining resonance scattering processes in soliton models there is no alternative to 
treating simultaneously the small amplitude fluctuations that correspond to asymptotic states 
and the collective rotations. At first sight this suggests double counting effects. However, 
Dirac constraints emerge that abandon Yukawa interactions, at least from the local and flavor 
symmetric part of the action. The Skyrme model description of this procedure is described in 
detail in Ref.~\cite{Walliser:2005pi} for the $\Theta^{+}$ channel in kaon-nucleon scattering. 
It combines rotations (collective motion) and vibrations (fluctuations) and is therefore 
called rotation-vibration-approach (RVA)\footnote{The omission of fluctuations, 
Eq.~(\ref{eq:coll1}) is called the rigid-rotator-approach (RRA) and finding the scattering 
data for the decoupled fluctuations goes by the label bound-state-approach 
(BSA)\cite{Callan:1985hy}.}. The key features of this approach are:
\begin{itemize}
\item[1)]
The field equations are solved in two steps, first for the soliton and second for the 
(constraint) fluctuations. These are the two leading orders in the $1/N_C$ counting. 
Consequently the above mentioned inconsistency with PCAC is not an issue for the RVA.
\item[2)]
The interaction between the collective models and the constrained fluctuations can be 
written as a separable potential that is similar to a resonance exchange. Reaction
matrix theory can be applied to this potential to compute the exchange phase shift
$\delta_E=\delta_E(N_C)$. The $N_C$ dependence originates from the matrix elements of
the collective coordinate operators in the separable potential. As indicated after 
Eq.~(\ref{ch6:hamrigid1}) they need to be taken between elements of representations
that vary with $N_C$.
\item[3)]
Up to a small (a few MeV) pole shift, $\delta_E$ passes through $\frac{\pi}{2}$ exactly 
at the momentum given by the mass difference between the $\Theta^{+}$ and the nucleon  
predicted by the collective coordinate Hamiltonian, Eq.~(\ref{ch6:hamrigid1}). Modulo 
flavor symmetry breaking effects, this difference decreases by a factor of two when 
going from $N_C=3$ to $N_C\to\infty$. 
\item[4)]
The reaction matrix formalism also derives a width function from the separable potential.
Up to flavor symmetry breaking this width function contains the matrix element of 
a \underline{single} collective coordinate operator between the nucleon and the 
$\Theta^+$. It is therefore impossible that substantial cancellations between matrix 
elements of several operators occur as was argued in Ref. \cite{Diakonov:1997mm}.
\item[5)]
The BSA gives exact results in the $N_C\to\infty$ limit. It is possible to modify
the BSA such that the scattering fluctuations are orthogonal to the rotations
described by the flavor rotations in Eq.~(\ref{eq:coll1}). For both BSA versions 
scattering phase shifts can be computed and their difference is the resonance phase 
shift $\delta_R$.
\end{itemize}
The litmus test for the RVA is then whether or not
$\lim_{N_C\to\infty}\delta_E(N_C)\stackrel{?}{=}\delta_R$.
The left panel of Fig.~\ref{fig:delta} impressively shows that this identity holds
over the whole range of relevant kaon momenta.

There are two important conclusions from this $\Theta^{+}$ analysis in the Skyrme model: 
First, the pentaquark resonance (state) is unavoidably predicted, it is not an artifact 
of the rigid rotator approach. Second, the large $N_C$ predictions from this model calculation
are based on the hedgehog structure of the soliton rather than on the particular model 
Lagrangian. Hence it is cogent that these predictions generalize to all chiral soliton 
models, thereby challenging results from the RRA reported in 
Refs.~\cite{Diakonov:1997mm,Ellis:2004uz}.

\section{Heavy baryons}
\label{heavy}

When turning to heavy baryons, not only those with exotic quantum numbers, consistency with 
the heavy quark spin/flavor symmetry~\cite{Eichten:1980mw,Neubert:1993mb} is the major 
requirement to model building. Essentially it states that for hadrons containing heavy 
quarks, spin and flavor effects are suppressed by inverse powers of the heavy quark mass. 
In turn this implies for effective mesons theories that heavy pseudoscalar and vector 
mesons are on equal footing\footnote{For an infinitely heavy quark, they are elements 
of a single multiplet~\cite{Schechter:1994ip}.}. This particularly implies that baryons 
with a heavy quark cannot be constructed as solitons within models solely based on the 
(pseudoscalar) chiral field.

A suitable approach is to include vector ($V^\mu$) and pseudoscalar ($P$) fields with 
masses $M^\ast$ and $M$, respectively \cite{Schechter:1995vr}. These fields represent 
mesons containing a single heavy quark and by themselves are three component vectors 
in flavor space that couple to the chiral field. When $M^\ast$ and $M$ approach infinity, 
$V^\mu$ and $P$ combine to a single multiplet whose dynamics is governed by the heavy quark 
spin/flavor symmetry. In the baryon sector the hedgehog configuration, Eq.~(\ref{eq:hedgehog}) 
induces an attractive potential for $V_\mu$ and $P$ such that bound state solutions with 
energies $|\omega|<M$ emerge. Many approaches based on this picture have been reported 
earlier~\cite{heavy}.

To leading order in the $1/N_C$ counting, these energies are the mass differences 
between baryons with a single heavy quark and the nucleon. More importantly, since the 
hedgehog dwells in the isospin subspace of the three dimensional flavor space, so does 
the two-component bound state wave function of the heavy meson fields \cite{Schechter:1995vr}: 
$$
P_{\rm b.s.}\propto 
{\rm e}^{-{\rm i}\omega t}\begin{pmatrix}\psi_p(\vec{x}\,)\cr 0\end{pmatrix}
\qquad {\rm and}\qquad
V^\mu_{\rm b.s.}\propto 
{\rm e}^{-{\rm i}\omega t}\begin{pmatrix}\psi^\mu_p(\vec{x}\,)\cr 0\end{pmatrix}\,.
$$
As in Eq.~(\ref{eq:coll1}) collective coordinate quantization approximates the full
field by the classical field rotating in flavor space: $P(\vec{x},t)=A(t)P_{\rm b.s}$
and $V^\mu(\vec{x},t)=A(t)V^\mu_{\rm b.s.}$. With 
$$
\dot{P}=-P{\rm i}{\rm e}^{-{\rm i}\omega t}A(t)
\left(\omega +\frac{1}{2}\sum_{a=1}^8\Omega_a\lambda_a\right)P_{\rm b.s.}
=-{\rm i}{\rm e}^{-{\rm i}\omega t}A(t)
\left(\omega+\frac{1}{2\sqrt{3}}\Omega_8
+\frac{1}{2}\sum_{a=1}^7\Omega_a\lambda_a\right)P_{\rm b.s.}
$$ 
and similarly for $V^\mu(\vec{x},t)$,
it is obvious that the Lagrangian contains contributions linear in $\Omega_8$. Appropriate
normalization \cite{Harada:1997we} of the bound state wave function, substitution into the 
Lagrange density and integrating over space provides the collective coordinate part of 
the Lagrange function from the heavy fields \cite{Blanckenberg:2015dsa}
\begin{equation}
L_h(\Omega_a)=-\omega\chi^\dagger \chi 
+\frac{1}{2\sqrt{3}}\chi^\dagger\Omega_8 \chi
+\frac{\rho}{2}\chi^\dagger\left(\vec{\Omega}\cdot\vec{\tau}\right)\chi\,.
\label{eq:hcc2}
\end{equation}
Here $\chi$ is the Fourier amplitude of the bound state wave function. Upon quantization 
$\chi^\dagger\chi$ becomes the number operator for the bound state. Furthermore $\rho$ is
a functional of all profile functions and describes the hyperfine splitting among the heavy 
baryons (see Eq.~(\ref{eq:master}) below). To linear order the angular velocities 
$\Omega_4,\ldots,\Omega_7$ do not contribute because the bound state wave function dwells 
in the isospin subspace.

\begin{figure}
\centerline{
\includegraphics[width=10cm,height=4cm]{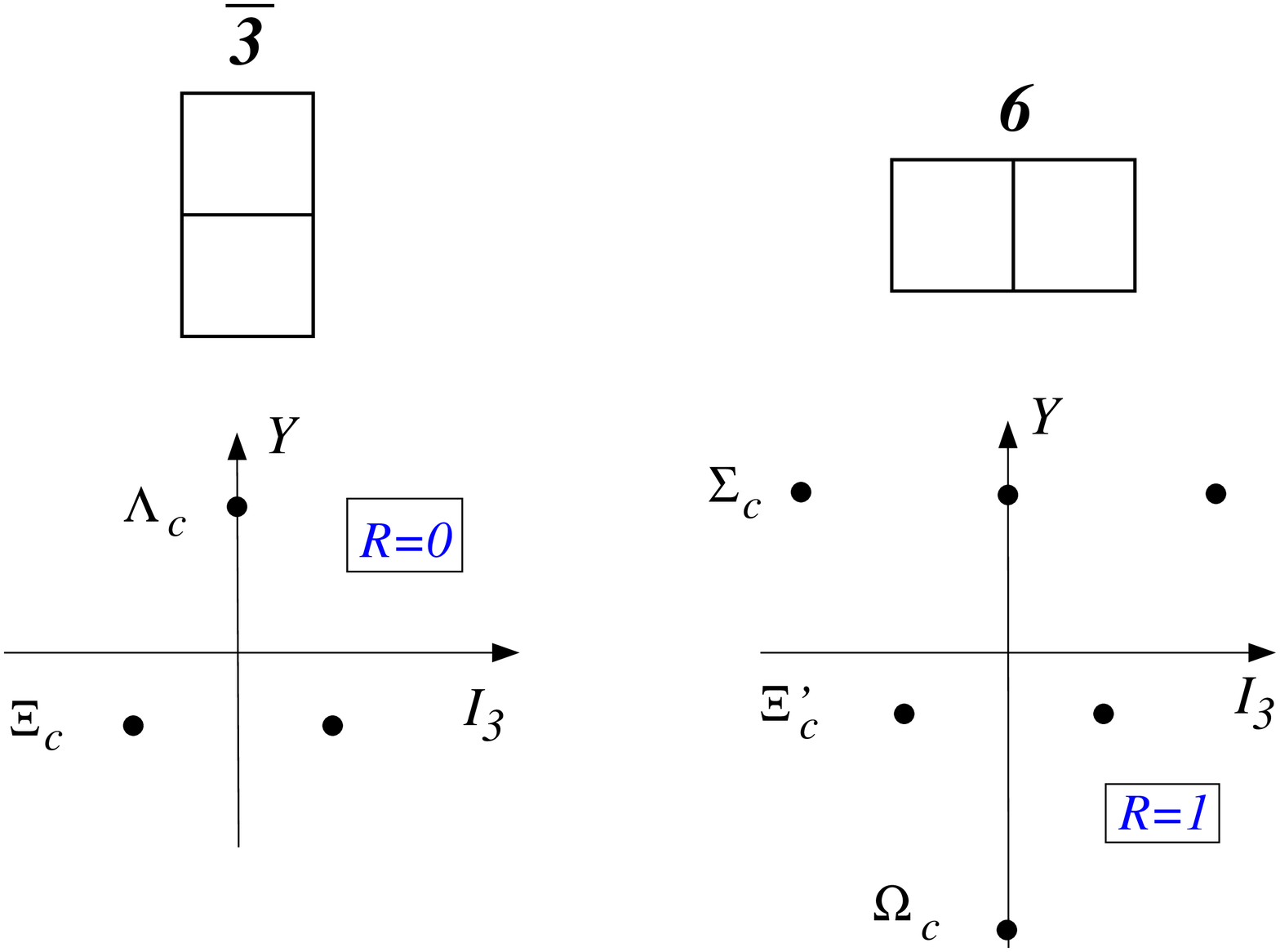}}
\smallskip
\caption{\label{fig:oheavy} $SU(3)$ representations for the diquark content 
of ordinary heavy baryons.}
\end{figure}

The main result to be extracted from Eq.~(\ref{eq:hcc2}) is its contribution to the 
constraint (assuming $N_C=3$ from now on) \cite{Momen:1993ax,Blanckenberg:2015dsa}
\begin{equation}
Y_R=1-\frac{1}{3}\chi^\dagger\chi
\,\longrightarrow\,\frac{3-n_H}{3}\,,
\label{eq:YRheavy}\end{equation}
where $n_H=+1$ when the bound state corresponds to a heavy quark ($0<\omega<M$) and 
$n_H=-1$ when the bound state corresponds to a heavy antiquark ($0>\omega>-M$). Hence for 
a bound heavy quark the lowest lying $SU(3)$ representations are the anti-triplet 
\parbox[l]{3mm}{\vskip-1.0mm\includegraphics[width=2mm,height=4mm]{adjoint.eps}} 
and the sextet 
\parbox[l]{5mm}{\vskip-0.8mm\includegraphics[width=4mm,height=2mm]{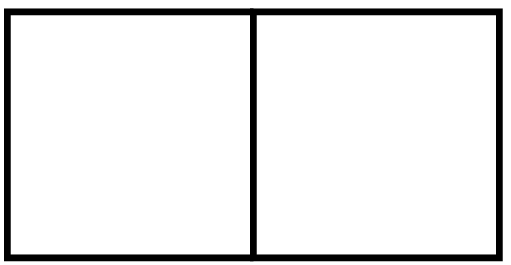}}. 
Obviously these representations describe the $uds$-diquark content of the heavy baryons. 
The particle content of the representations is displayed in Fig.~\ref{fig:oheavy}.
Of course, none-occupation ($n_H=0$) of the heavy meson bound state falls back to the 
ordinary baryons, as in Fig.~\ref{fig:ground}.

Quantizing the canonical momentum $\vec{R}=-\frac{\partial L}{\partial\vec{\Omega}}$
yields the {\it intermediate} spin quantum number $R$. The hedgehog 
structure causes $R$ to equal the isospin of the zero strangeness element in a given 
representation. That is, $R=0$ for the anti-triplet and $R=1$ for sextet. Adding
the intrinsic spin of the heavy meson bound state gives the total baryon spin
$\vec{J}=-\vec{R}-\chi^\dagger(\vec{\tau}/2)\chi$. In total the mass formula reads
\cite{Blanckenberg:2015dsa}
\begin{equation}
M-E_{\rm cl}=\left(\frac{1}{\alpha^2}-\frac{1}{\beta^2}\right)\frac{R(R+1)}{2}
+\frac{\epsilon}{2\beta^2}
-\frac{3}{8\beta^2}\left(1-\frac{n_H}{3}\right)^2
+|\omega|n_H+\frac{\rho}{2\alpha^2}\left[J(J+1)-R(R+1)\right]n_H\,,
\label{eq:master}
\end{equation}
where $\epsilon$ is the eigenvalue obtained from diagonalizing flavor symmetry breaking: 
$\left\{C_2[SU(3)]+2\beta H_{\rm SB}\right\}\Psi=\epsilon \Psi$ for prescribed $I$, $R$ and 
$Y_R$ \cite{Yabu:1987hm}. Factorizing $2\beta H_{\rm SB}=\lambda\Gamma_{\rm SB}$, with 
$\Gamma_{\rm SB}$ a fixed value determined from the profile functions, we  consider
$\lambda=\frac{2m_s}{m_u+m_d}$ a tuneable parameter to eventually study non-linear 
effects of flavor symmetry breaking. Model results for the baryon spectrum computed
from Eq.~(\ref{eq:master}) have been reported in Ref. \cite{Blanckenberg:2015dsa} and
there is no need to repeat them here. However, it is important to stress that, both the 
constraint on $Y_R$, that selects the relevant $SU(3)$ representation(s), and the form of 
the hyperfine splitting (last term in Eq.~(\ref{eq:master})) inherently arise from the model 
calculation once compulsory extensions of the mean field treatment are implemented. This 
makes fully obsolete the adjustment of the classical action (using external specifications)
to accommodate the heavy sector as in Ref.~\cite{Yang:2016qdz}. Such an accommodation causes 
the moments of inertia ($\alpha^2$ and $\beta^2$) to vary with $n_H$ and may even require to 
modify the leading order (in $1/N_C$) field equation. Stated otherwise, without an extension 
of the mean field approach the heavy and light sectors cannot be connected consistently. 

\begin{figure}
\centerline{
\includegraphics[width=10cm,height=3cm]{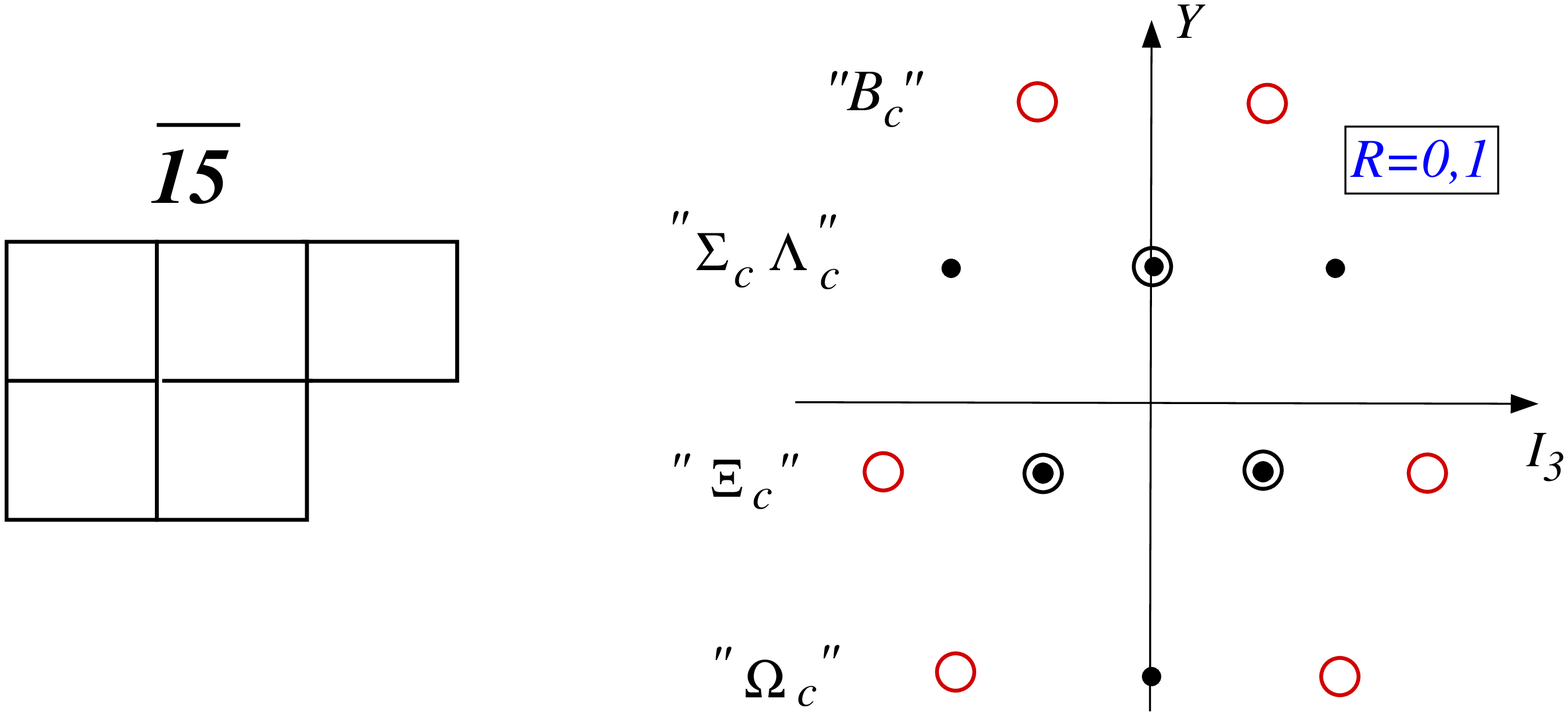}}
\smallskip
\caption{\label{fig:oheavy1}(Color online) Young tableaux and particle content of the 
anti-decapentaplet ($\mathbf{\overline{15}}$). Black circles denote ordinary diquarks whose 
quantum numbers relate to two quark states. Double circles indicate two states with different 
$R$ quantum numbers. Red circles denote exotic states that cannot be built from two quarks.}
\end{figure}

Similar to the anti-decuplet ($\mathbf{\overline{10}}$) in the non-heavy sector the next to 
lowest representation, the anti-decapentaplet ($\mathbf{\overline{15}}$), shown in 
Fig.~\ref{fig:oheavy1}, plays a twofold role for the spectrum of heavy baryons. 
First, the $\mathbf{\overline{15}}$ contains diquarks with the same (observable)
quantum numbers as the members of the $\mathbf{\overline{3}}$ and/or $\mathbf{6}$. When
diagonalizing flavor symmetry breaking these member states mix to build the eigenfunctions
$\Psi$. Second, the $\mathbf{\overline{15}}$ contains {\it exotic} members (red circles 
in Fig.~\ref{fig:oheavy1}) that represent composites of three quarks and an anti-quark
from the $uds$ sector.
\begin{figure}
\centerline{
\includegraphics[width=5cm,height=4.5cm]{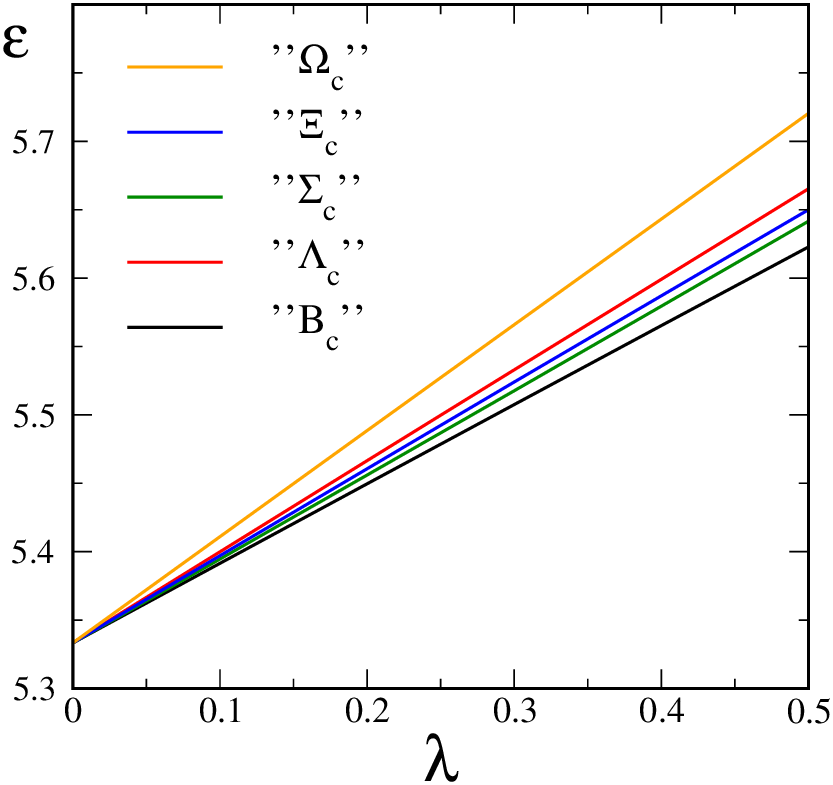}\hspace{1cm}
\includegraphics[width=8cm,height=4.5cm]{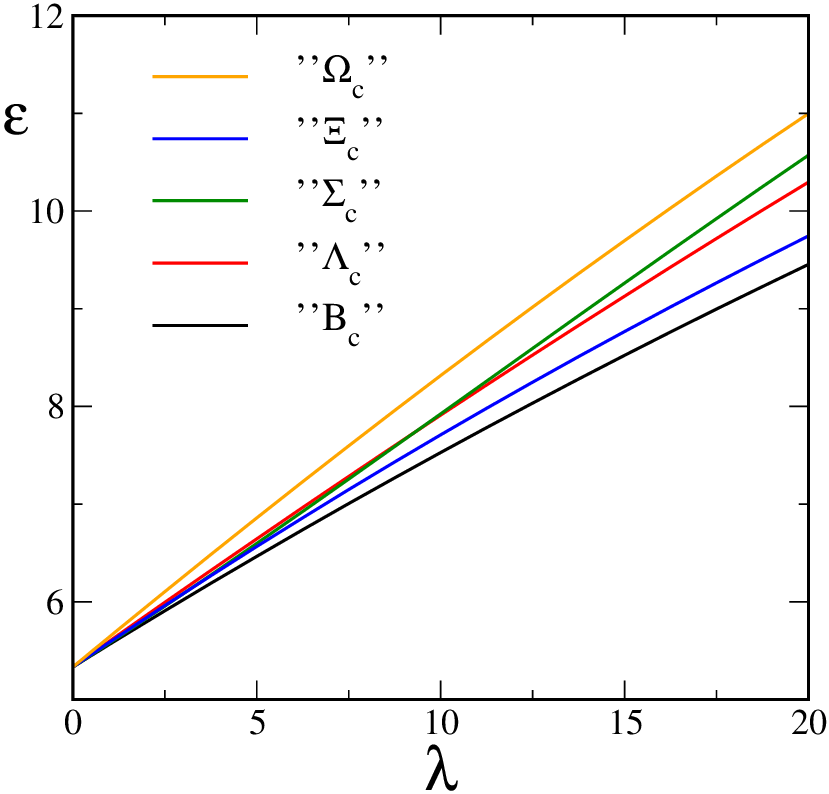}}
\caption{\label{fig:sbr15}(Color online) Flavor symmetry breaking eigenvalues for states 
that at $\lambda=0$ are pure $\mathbf{\overline{15}}$ states, {\it cf.} Fig.~\ref{fig:oheavy1}. 
The parameter $\lambda$ measures the strength of symmetry breaking.}
\end{figure}
In Fig.~\ref{fig:sbr15} we show the flavor symmetry breaking eigenvalue of these states as a 
function of the strength of the symmetry breaking. We actually observe that the order of the 
eigenvalues for $\Xi$ and $\Sigma$ type states is reversed as the strength assumes its 
empirical value, $\lambda\sim20$. Also the eigenvalue for $B_c$, the candidate for the lowest
charmed exotic baryon, significantly deviates from a straight line. These findings doubt 
the truncation of the perturbation expansion at the first order \cite{Kim:2017jpx}
when describing the spectrum of these states. The fact that eigenvalues grow less than 
linear also leads to a lower prediction for the masses of the exotic heavy baryons. Using 
the parameters as determined in Ref. \cite{Blanckenberg:2015dsa} predicts
$$
M(B_c)-M(\Lambda_c)\approx 495{\rm MeV}\,,\quad 
M(''\Xi_c'')-M(\Lambda_c)\approx 530{\rm MeV}\,,\quad
M(''\Omega_c'')-M(\Lambda_c)\approx 670{\rm MeV}
$$
for the mass differences of the charmed exotic baryons.

\begin{figure}
\centerline{
\includegraphics[width=10cm,height=2.5cm]{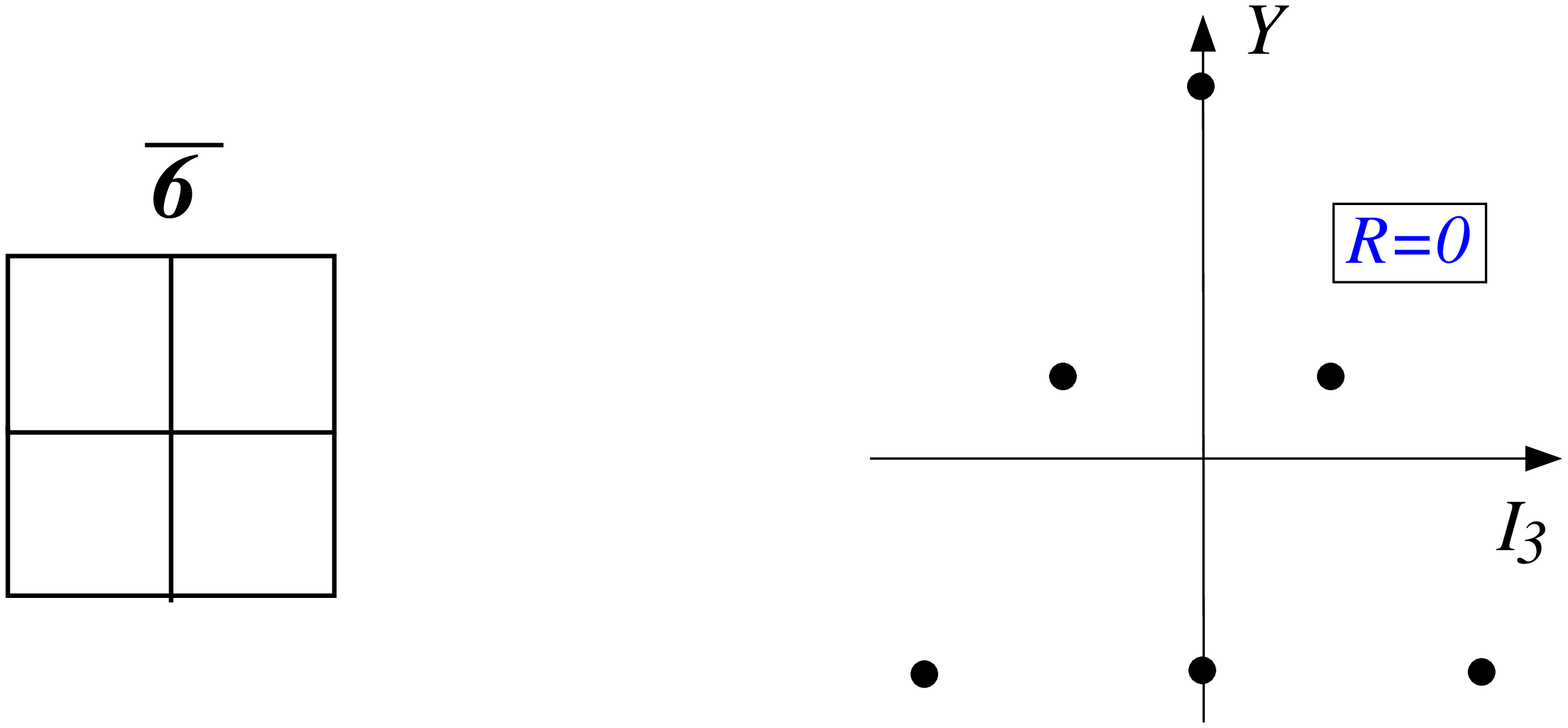}}
\smallskip
\caption{\label{fig:bar6}Young tableaux and flavor content of the 
anti-sextet representation.}
\end{figure}

Finally let us have a brief look at pentaquarks with a heavy antiquark. They are constructed 
from heavy meson bound states with negative energy eigenvalues. The resulting binding energy
is significantly less than that of ordinary bound states and may even be unbound in the 
charm sector~\cite{Schechter:1995vr}. Since negative energy eigenvalues correspond to
$n_H=-1$ and $Y_R=\frac{4}{3}$, the lowest dimensional representation available 
is the anti-sextet ($\mathbf{\overline{6}}$) representing four (light) quarks as 
shown in Fig.~\ref{fig:bar6}.
This representation has $R=0$ and thus the states must be compared to those from the 
anti-triplet. The collective coordinate quantization induces a mass difference 
$M(\mathbf{\overline{6}})-M(\mathbf{\overline{3}})=\frac{2}{\beta^2}+\mbox{sym. breaking}
\sim 300{\rm MeV}$ which is much larger than the typical binding energy. Hence this
model estimate suggests that such baryons most probably do not exist.

\section{Summary}
\label{summary}

We have reviewed two examples for the description of exotic baryons in chiral soliton models. 
The main purpose of this endeavor has been to reflect on conceptual issues rather than any
accurate reproduction of empirical data by tuning parameters. Though most of the results are 
based on the Skyrme model treatment of the chiral field, they generalize to any chiral model 
with a hedgehog soliton. 

Exotic baryons naturally emerge in chiral soliton models as elements in higher dimensional
representations when quantizing the three light flavor degrees of freedom. Then these baryons 
are flavor rotational excitations so that their excitation energy may be significantly lower 
than expected from the respective constituent quark masses. Of course, there is reason to 
assume this soliton model prediction could be a mere artifact of the collective coordinate 
approximation. However, a careful analysis of the $\Theta^{+}$ pentaquark reveals that this 
is not the case and the $\Theta^{+}$ indeed shows up as a resonance in the Skyrme model analysis 
of kaon nucleon scattering. As a by-product of this analysis we have seen that any estimate of 
the hadronic width of (exotic) baryon resonances must include the subtle interplay of collective 
and vibrational modes. Obviously such estimates are more complicated than just generalizing the 
Goldberger-Treiman relation.

In the context of baryons with a heavy quark (or antiquark) we have seen that the heavy meson bound 
state wave function, when properly included in the collective coordinate approximation, automatically 
selects the pertinent light flavor representation. This concerns both types of exotic heavy baryons,
those with a heavy antiquark as well as those with a light antiquark. The presented approach does not
account for a {\it back-reaction} of the heavy meson bound state on the chiral field. Since the bound
state couples at $\mathcal{O}(N_C^0)$ this will not affect the soliton itself because it obeys a field
equation that arises from the $\mathcal{O}(N_C)$ part of the action.

The main conclusion is that the description of exotic baryons within chiral soliton models is not 
complete on the mean field level. It would require specifications from the outside which are not 
guaranteed to be consistent within the model itself. As examples thereof we have argued 
against the fabrication of a Yukawa coupling and the adjustment of the constraint 
that selects the flavor $SU(3)$ representation. On the contrary, the proper incorporation of (harmonic) 
fluctuations, which goes beyond the mean field approach, produces these specifications in a self-contained 
manner. Hence such fluctuations are essential to understand the structure and properties of exotic baryons 
in chiral soliton models.

\section*{Acknowledgement}
Some of the results presented here were obtained in collaborations with J. Schechter, H. Walliser and 
J. P. Blanckenberg whose contributions are gratefully acknowledged. The author would also like to 
thank the organizers to make the {\it International Conference on New Frontier in Physics} a worthwhile
event. The project is supported in part by the National Research Foundation of South Africa (NRF)
by grant~109497.


\begin{thebibliography}{99}
\bibitem{Walliser:2005pi}
  H.~Walliser and H.~Weigel,
  Eur.\ Phys.\ J.\ A {\bf 26} (2005) 361.

\bibitem{Weigel:2007yx}
  H.~Weigel,
  Phys.\ Rev.\ D {\bf 75} (2007) 114018.

\bibitem{Blanckenberg:2015dsa}
  J.~P.~Blanckenberg and H.~Weigel,
  Phys.\ Lett.\ B {\bf 750} (2015) 230.

\bibitem{Weigel:2008zz}
  H.~Weigel,
  Lect.\ Notes Phys.\  {\bf 743} (2008) 1.

\bibitem{Diakonov:1997mm}
  D.~Diakonov, V.~Petrov and M.~V.~Polyakov,
  Z.\ Phys.\ A {\bf 359} (1997) 305.

\bibitem{Ellis:2004uz}
  J.~R.~Ellis, M.~Karliner and M.~Praszalowicz,
  JHEP {\bf 0405} (2004) 002.

\bibitem{Kim:2017khv}
  H.~C.~Kim, M.~V.~Polyakov, M.~Praszalowicz and G.~S.~Yang,
  Phys.\ Rev.\ D {\bf 96} (2017)  094021;
   Erratum: [Phys.\ Rev.\ D {\bf 97} (2018)  039901].

\bibitem{Praszalowicz:2018upb}
  M.~Praszalowicz,
  Eur.\ Phys.\ J.\ C {\bf 78} (2018) 690.


\bibitem{Eichten:1980mw}
  E.~Eichten and F.~Feinberg,
  Phys.\ Rev.\ D {\bf 23} (1981) 2724;~~
  M.~A.~Shifman and M.~B.~Voloshin,
  Sov.\ J.\ Nucl.\ Phys.\  {\bf 45} (1987) 292 [Yad.\ Fiz.\  {\bf 45} (1987) 463];~~
  N.~Isgur and M.~B.~Wise,
  Phys.\ Lett.\ B {\bf 237} (1990) 527;~~
  H.~Georgi,
  Phys.\ Lett.\ B {\bf 240} (1990) 447.

\bibitem{Neubert:1993mb}
  M.~Neubert,
  Phys.\ Rept.\  {\bf 245} (1994) 259.

\bibitem{Yang:2016qdz}
  G.~S.~Yang, H.~C.~Kim, M.~V.~Polyakov and M.~Praszałowicz,
  Phys.\ Rev.\ D {\bf 94} (2016) 071502

\bibitem{Kim:2018cxv}
  H.~C.~Kim,
  J.\ Korean Phys.\ Soc.\  {\bf 73} (2018) 165.

\bibitem{Callan:1985hy}
  C.~G.~Callan, Jr. and I.~R.~Klebanov,
  Nucl.\ Phys.\ B {\bf 262} (1985) 365.

\bibitem{Witten:1979kh}
E.~Witten {\em Nucl. Phys.} {\bf B160} (1979) 57.

\bibitem{Skyrme:1961vq}
  T.~H.~R.~Skyrme,
  Proc.\ Roy.\ Soc.\ Lond.\ A {\bf 260} (1961) 127;
  Int.\ J.\ Mod.\ Phys.\ A {\bf 3} (1988) 2745 {\it (article
  reproduced by I. Aitchison.)}.

\bibitem{Adkins:1983ya}
  G.~S.~Adkins, C.~R.~Nappi and E.~Witten,
  Nucl.\ Phys.\ B {\bf 228} (1983) 552.

\bibitem{Jain:1989kn}
  P.~Jain, R.~Johnson, N.~W.~Park, J.~Schechter and H.~Weigel,
  Phys.\ Rev.\ D {\bf 40} (1989) 855.

\bibitem{Wi83}
E.~Witten {\em Nucl. Phys.} {\bf B223} (1983) 422, 433.

\bibitem{Meier:1996ng}
  F.~Meier and H.~Walliser,
  Phys.\ Rept.\  {\bf 289} (1997) 383.

\bibitem{Guadagnini:1983uv}
  E.~Guadagnini,
  Nucl.\ Phys.\ B {\bf 236} (1984) 35.

\bibitem{Jain:1984gp}
  S.~Jain and S.~R.~Wadia,
  Nucl.\ Phys.\ B {\bf 258} (1985) 713.

\bibitem{Alkofer:1994ph}
  R.~Alkofer, H.~Reinhardt and H.~Weigel,
  Phys.\ Rept.\  {\bf 265} (1996) 139.

\bibitem{Biedenharn:1984su}
  L.~C.~Biedenharn and Y.~Dothan,
  ``Monopolar Harmonics in $SU(3)_f$ as Eigenstates of the Skyrme-Witten Model for Baryons'',
  in E. Gotsman, G. Tauber (Eds.), ''From SU(3) To Gravity'', (1985) 15.

\bibitem{Park:1989by}
  N.~W.~Park, J.~Schechter and H.~Weigel,
  Phys.\ Lett.\ B {\bf 228} (1989) 420.

\bibitem{Yabu:1987hm}
  H.~Yabu and K.~Ando,
  Nucl.\ Phys.\ B {\bf 301} (1988) 601.

\bibitem{Dorey:1994fk}
  N.~Dorey {\it et al.} in Ref. \cite{width}

\bibitem{Newton:1982qc}
R. G. Newton, {\it Scattering Theory of Waves and Particles}
Springer, New York (1982), chap. 11.2.2.

\bibitem{Schwesinger:1988af}
  B.~Schwesinger, H.~Weigel, G.~Holzwarth and A.~Hayashi,
  Phys.\ Rept.\  {\bf 173} (1989) 173.

\bibitem{width}
  M.~Uehara,
  Prog.\ Theor.\ Phys.\  {\bf 75} (1986) 212
  Erratum: [Prog.\ Theor.\ Phys.\  {\bf 75} (1986) 464];
  Prog.\ Theor.\ Phys.\  {\bf 78} (1987) 984;~~
  S.~Saito,
  Prog.\ Theor.\ Phys.\  {\bf 78} (1987) 746;~~
  G.~Holzwarth, A.~Hayashi and B.~Schwesinger,
  Phys.\ Lett.\ B {\bf 191} (1987) 27;~~
  H.~Verschelde,
  Phys.\ Lett.\ B {\bf 209} (1988) 34;~~
  D.~Diakonov, V.~Y.~Petrov and P.~V.~Pobylitsa,
  Phys.\ Lett.\ B {\bf 205} (1988) 372;~~
  G.~Holzwarth,
  Phys.\ Lett.\ B {\bf 241} (1990) 165;~~
  G.~Holzwarth, G.~Pari and B.~K.~Jennings,
  Nucl.\ Phys.\ A {\bf 515} (1990) 665;~~
  A.~Hayashi, S.~Saito and M.~Uehara,
  Phys.\ Lett.\ B {\bf 246} (1990) 15;
  Phys.\ Rev.\ D {\bf 43} (1991) 1520;
  Prog.\ Theor.\ Phys.\ Suppl.\  {\bf 109} (1992) 45;~~
  N.~Dorey, J.~Hughes and M.~P.~Mattis,
  Phys.\ Rev.\ D {\bf 50} (1994) 5816.

\bibitem{Schechter:1994ip}
  J.~Schechter and A.~Subbaraman,
  Phys.\ Rev.\ D {\bf 51} (1995) 2311.

\bibitem{Schechter:1995vr}
  J.~Schechter, A.~Subbaraman, S.~Vaidya and H.~Weigel,
  Nucl.\ Phys.\ A {\bf 590} (1995) 655;
   Erratum: [Nucl.\ Phys.\ A {\bf 598} (1996) 583].

\bibitem{heavy}
E.~E.~Jenkins and A.~V.~Manohar,
  Phys.\ Lett.\ B {\bf 294} (1992) 273;~~
Z.~Guralnik, M.~E.~Luke and A.~V.~Manohar,
  Nucl.\ Phys.\ B {\bf 390} (1993) 474;~~
E.~E.~Jenkins, A.~V.~Manohar and M.~B.~Wise,
  Nucl.\ Phys.\ B {\bf 396} (1993) 38;~~
D.~P.~Min, Y.~s.~Oh, B.~Y.~Park and M.~Rho,
  hep-ph/9209275;~~
H.~K.~Lee, M.~A.~Nowak, M.~Rho and I.~Zahed,
  Annals Phys.\  {\bf 227} (1993) 175;~~
M.~A.~Nowak, I.~Zahed and M.~Rho,
  Phys.\ Lett.\ B {\bf 303} (1993) 130;~~
D.~P.~Min, Y.~s.~Oh, B.~Y.~Park and M.~Rho,
  Int.\ J.\ Mod.\ Phys.\ E {\bf 4} (1995) 47;~~
K.~S.~Gupta, M.~A.~Momen, J.~Schechter and A.~Subbaraman,
  Phys.\ Rev.\ D {\bf 47} (1993) R4835;~~
A.~Momen, J.~Schechter and A.~Subbaraman,
  Phys.\ Rev.\ D {\bf 49} (1994) 5970;~~
Y.~S.~Oh, B.~Y.~Park and D.~P.~Min,
  Phys.\ Rev.\ D {\bf 49} (1994) 4649;
  Phys.\ Rev.\ D {\bf 50} (1994) 3350;~~
J.~Schechter and A.~Subbaraman,
  Phys.\ Rev.\ D {\bf 51} (1995) 2311;~~
Y.~S.~Oh and B.~Y.~Park,
  Phys.\ Rev.\ D {\bf 51} (1995) 5016.

\bibitem{Harada:1997we}
  M.~Harada, A.~Qamar, F.~Sannino, J.~Schechter and H.~Weigel,
  Nucl.\ Phys.\ A {\bf 625} (1997) 789.

\bibitem{Momen:1993ax}
A.~Momen {\it et al.} in Ref.~\cite{heavy}

\bibitem{Kim:2017jpx}
  H.~C.~Kim, M.~V.~Polyakov and M.~Praszałowicz,
  Phys.\ Rev.\ D {\bf 96} (2017) 014009;
   Addendum: [Phys.\ Rev.\ D {\bf 96} (2017) 039902].

\end{thebibliography}
\end{document}